\documentstyle[preprint,aps,epsf]{revtex}
\newcommand{\iunit}{{\rm i}}
\newcommand{\halb}{\frac{1}{2}}
\newcommand{\onehalf}{\frac{1}{2}}

\title{Angular Correlations in Internal Pair Conversion of 
Aligned Heavy Nuclei}
\author{C.~R.~Hofmann\thanks{E-Mail: hofmann@ptprs8.phy.tu-dresden.de},\
         G.~Soff \\
         {\it Institut f\"ur Theoretische Physik, TU Dresden} \\
         {\it D-01062 Dresden} \\[0.5cm]
         J.~Reinhardt\thanks{E-Mail: jr@th.physik.uni-frankfurt.de},\ 
         W.~Greiner \\
         {\it Institut f\"ur Theoretische Physik, J.~W.~Goethe Universit\"at}\\
         {\it D-60054 Frankfurt am Main}}
\begin{document}
\maketitle
\newpage
\begin{abstract}
We calculate the spatial correlation of electrons and positrons emitted
by internal pair conversion of Coulomb excited nuclei in 
heavy ion collisions. The alignment or polarization of
the nucleus results in an anisotropic emission of the electron-positron
pairs which is closely related to the anisotropic 
emission of $\gamma$-rays. However, the angular correlation 
in the case of internal pair conversion exhibits 
diverse patterns. This
might be relevant when investigating atomic processes in 
heavy-ion collisions performed at the Coulomb barrier.
\end{abstract}
\newpage
%
%
%
%
%
\section{Introduction}
%
%
Heavy-ion collisions at energies in the vicinity of the nuclear 
Coulomb barrier
lead to an alignment of the colliding nuclei. This implies 
that the magnetic
substates are no longer equally populated. To describe deexcitation
processes following heavy-ion collisions such as $\gamma$-ray emission or
internal conversion, we have to account for this 
specific population by weighting the transition
matrix elements with the occupation probability of, rather than just 
averaging over the decaying substates.

The population of the various nuclear substates 
is incorporated in the formalism by introducing
the density matrix of the excited quantum system or, in the case
of rotational symmetry of the problem, by a set of statistical tensors 
which obey the
same transformation law as the spherical harmonics. This concept enables us
to treat the polarization or alignment of excited nuclei appropriately.
First calculations of the angular correlation of electrons and positrons
emitted in internal pair conversion taking into account the alignment
of nuclei were accomplished by Goldring, Rose and Warburton 
\cite{goldring,rose:63,warburton}. These calculations were performed within
Born approximation, neglecting the influence of the nuclear charge on the
outgoing electron and positron. But for 
internal pair conversion (IPC) of highly charged nuclei the
Born approximation is not justified as can be verified by the corresponding
positron spectra \cite{schlueter:78,schlueter:81,soff:81}.

Therefore we reconsider in the following the internal pair conversion 
of heavy nuclei which
are aligned, e.g., by Coulomb excitation or transfer reactions. 
We determine the angular correlation 
of the emitted electron and positron with respect 
to a reference axis in space. As 
already known for the angular correlation of $\gamma$ rays, the problem will
be simplified if we choose a coordinate system in which the density matrix is
diagonal. The statistical tensors depend as well on the choice of the
coordinate system. If the entries of the statistical tensors 
are given in a specific coordinate system, we are able
to calculate the angular correlation with respect to the $z$ axis of
this system.

The occupation probabilities of the magnetic substates 
caused by Coulomb excitation can be calculated with, e.g.,
the COULEX code of K.~Alder and A.~Winther \cite{alder:winther:1}.
However, one should take into account the change of
population by electromagnetic transitions from 
higher lying states. Special attention should be paid
to a proper choice of the coordinate system when dealing with the COULEX code
\cite{alder:winther:1,alder:winther:2}. 
For pure Coulomb excitation we will assume the $z$ axis to point
along the asymptotic target recoil axis. With respect to this axis
the excited nuclei may exhibit prolate or oblate alignment.
%
%
\section{Density matrix and statistical tensors}
%
%
The density matrix -- and for spherical symmetry the set of statistical
tensors  -- is the appropriate tool for including statistical
properties such as 
occupation probabilities of quantum mechanical states 
into the calculations.
Here we briefly summarize the essential properties of the density matrix 
and subsequently turn
to the concept of the statistical tensors, which obey the same
transformation law as irreducible tensors. For the density matrix 
$\rho_{M_i M'_i}(J_i)$ of dimension $(2J_i+1)\times(2J_i+1)$ we note:
\begin{enumerate}
\item
The density matrix is hermitian: 
$$ \rho^{*}_{M'_i M_i}( J_i ) = \rho_{M_i M'_i}( J_i ) \quad , $$
\item
The trace of the density matrix equals one:
$$ 
{{\rm Tr} \{ \rho \} } = 1 \Leftrightarrow
\sum\limits_{M_i} \rho_{M_i M_i}( J_i ) = 1 \quad , $$
\item
${\rm Tr} \{ \rho^2 \} \leq 1 \quad,\quad \mbox{and} \quad 
{\rm Tr} \{ \rho^2 \} = 1 \Leftrightarrow 
\mbox{the system is in a pure state.}$
\end{enumerate}

We define the statistical tensors $\hat\rho^{[n]}_{\nu}$ as irreducible tensors 
of rank $n$ with $\nu=-n,\ldots,n$:
\begin{equation}
\hat\rho^{[n]}_{\nu}(J_i) = \sum_{M_i,M'_i} (-1)^{J_i-M_i'} \sqrt{2n+1}
\left( \begin{array}{ccc} J_i & J_i & n \\ M_i & M'_i & -\nu
\end{array}\right) \rho_{M_i M'_i}( J_i )
\quad .
\label{dichtetensor-hin}
\end{equation}
The argument $J_i$ reminds us that $n$ is related to the angular momentum
of the magnetic substates by $0 \leq n \leq 2J_i$. 
From the normalization of the density matrix it follows
$\hat\rho^{[0]}_{0}(J_i) = 1/\sqrt{2J_i+1}$.

The density matrix has $(2J_i+1)^2$ independent components. To describe a
system by statistical tensors instead of the density matrix, we need
$2J_i+1$ density tensors of rank $n=0$ up to rank $n=2J_i$. Since
the density tensor of rank $n$ has $2n+1$ components we get again
$\sum_{n=0}^{2J_i}\left(\sum_{\nu=-n}^{n} 1 \right) = (2J_i+1)^2$ independent
components.
 
The statistical tensors transform under rotations according to 
\begin{equation}
\hat\rho^{[n]}_{\nu}(J_i) = \sum\limits_{\nu'}
{\cal D}^{[n]*}_{\nu' \nu}( \vec\alpha ) \, \hat\rho^{[n]}_{\nu'}(J_i)
\end{equation} 
where the Wigner rotation matrix of rank $n$ is denoted by ${\cal D}^{[n]}$ 
and the set of Euler angles by $\vec\alpha$. In defining the Euler angles
we follow Rose \cite{rose:57} and Eisenberg\&Greiner 
\cite{eisenberg:greiner:1}.
For systems with rotational symmetry it is thus more advantageous to 
employ the concept of the statistical
tensors when incorporating statistical statements concerning the system. 
The components of the statistical tensors are changed under rotations
and so are the occupation numbers of the magnetic substates.

From the set of $2J_i+1$ statistical tensors we obtain the density matrix by
utilizing the relation:
\begin{equation}
\rho_{M_i M_i'}( J_i ) = (-1)^{J_i-M'_i} \sum\limits_{n,\nu} \sqrt{2n+1}
\left(\begin{array}{ccc} J_i & J_i & n \\ M_i' & -M_i & -\nu \end{array}\right)
\hat\rho^{[n]}_{\nu}(J_i)
\label{dichtetensor-rueck}
\quad .
\end{equation}

For certain symmetries of the system we can reduce the independent components
of the statistical tensors.
Here we list the consequences for the statistical tensors in three special
cases which will become relevant for us:
\begin{enumerate}

\item

In the case of {\em axial symmetry} the density matrix is diagonal, its
diagonal components are just the probabilities for the occupation of the
corresponding magnetic substates $\rho_{M_i M_i} = p_{M_i}$:
$$
\hat\rho^{[n]}_{\nu}(J_i) =
\delta_{0\nu}\sum\limits_{M_i} (-1)^{J_i-M_i} p_{M_i}
\left(\begin{array}{ccc} J_i & J_i & n \\ M_i & -M_i & 0
\end{array}\right)
\quad .
$$
One can always choose a basis such that the density matrix is diagonal,
but in general this will not be a basis of wave functions with good
angular momentum.

\item

In the case of {\em spherical symmetry}, there is no direction singled 
out in space.
The density matrix is proportional to the identity matrix. The diagonal
elements are given by $\rho_{M_i M_i} = 1/(2J_i+1)$. All statistical tensors
vanish with exception of the tensor of rank $0$, i.e.,
$$
\hat\rho^{[n]}_{\nu}(J_i) = \delta_{0n}\,\delta_{0\nu} 
\frac{1}{\sqrt{2J_i+1}} 
\quad.
$$

\item

From Eq.~(\ref{dichtetensor-hin}) it can be shown that for alignment 
of the nuclear states, defined by $ p_{M_i} = p_{-M_i} $,
the statistical tensors of odd rank vanish.
\end{enumerate}
%
%
\section{Angular correlation of $\gamma$-rays}
%
%
Before we enter into the calculations concerning the angular correlation
in internal pair conversion we summarize some results already known
for in-beam $\gamma$-ray spectroscopy.
This will help us to interpret the angular correlation pattern in the case
of internal pair conversion. The angular correlation of photons emitted
after Coulomb excitation is 
essentially determined by the statistical tensors, i.e.,
by the occupation numbers of the magnetic substates of the decaying nucleus.
In choosing a reference axis for which the density matrix is diagonal, just
the 0th components of all statistical tensors survive and we obtain for the
transition probability the well-known relations:
\begin{equation}
\frac{{\rm d}P_\gamma}{{\rm d}\Omega}
= \frac{2\alpha\omega}{\sqrt{2J_i+1}} \left| V_\gamma^{(\tau)}(L) \right|^2
\sum\limits_{I\ {\rm even}} F_I( L \, L \, J_f \, J_i )
\; \hat\rho^{[I]}_{0}(J_i) \, P_I(\cos\vartheta)
\quad , 
\end{equation}
for a transition of parity $\tau={\rm E/M}$ and multipolarity $L$.
$V_\gamma^{(\tau)}(L)$ denotes the corresponding reduced matrix
element for the nuclear transition.
Here we employed the correlation coefficients 
\cite{biedenharn,schwalm,wollersheim}
\begin{eqnarray}
F_I( L \, L \, J_f \, J_i ) & = & (-)^{J_f+J_i -1} \,
\sqrt{2I+1} \, \sqrt{ 2J_i+1}
\nonumber
\\
& \times & 
(2L+1)
\left(\begin{array}{ccc} L & L & I \\ 1 & -1 & 0 \end{array}\right)
\left\{\begin{array}{ccc} L & L & I \\ J_i & J_i & J_f \end{array}\right\}
\label{ph-emission-20}
\; .
\end{eqnarray}
This results in an anisotropic emission of photons with respect to the
alignment axis. The number of minima of the angular distribution corresponds
to the multipolarity of the nuclear transition. 

In the case of spherical symmetry, the photon emission is isotropic
\begin{equation}
\frac{{\rm d}P_\gamma}{{\rm d}\Omega} = 
\frac{2\alpha\omega}{2J_i+1} \, \left| V_\gamma^{(\tau)}(L)\right|^2
\end{equation} 
or integrated over the solid angle $\Omega$:
\begin{equation}
P_\gamma = \frac{8\pi\alpha\omega}{2J_i+1} \, 
\left| V_\gamma^{(\tau)}(L)\right|^2
\quad .
\label{photon}
\end{equation}
%
%
\section{Transition probabilities for internal pair\protect\newline 
         conversion}
%
%
We turn now to the formulation of the triple correlation of the 
electron and positron direction with reference to a symmetry axis,
which is taken as quantization axis.

For a statistical ensemble of nuclei we write the transition probability
for internal pair conversion,
\begin{eqnarray}
   P_{e^{+}e^{-}}
& = &
   2\pi
   \sum\limits_{M_i , M_i' , M_f , \lambda , \lambda'}
   \int {\rm d}^{3}p \int {\rm d}^{3}p' \, \delta(\omega - W' - W) 
\nonumber
\\
& & 
   \times
   U_{\rm pl}\; \rho_{M_i M_i'} \; U^{*}_{\rm pl}
   \quad .
   \label{p}
\end{eqnarray}
where the density matrix $\rho_{M_i M'_i}$ represents the occupation of the
magnetic substates $|J_i M_i\rangle$. 
Here we assumed a nuclear transition from a initial state $|J_i M_i\rangle$
to the final state $|J_f M_f\rangle$ where the initial state is populated
according to the density matrix $\rho_{M_i M'_i}$. 
Since we do not require the density matrix to be diagonal the
summation extends over both, $M_i$ and $M'_i$. The $\delta$ function
ensures energy conservation: The transition energy $\omega$ is transfered
to the electron (total energy $W'$) and to the positron (total energy $W$).
The summation is taken over the spins and the momenta of the outgoing leptons.

The matrix element for internal pair conversion is written in lowest
order of $\alpha$ in the retarded form:
\begin{equation}
U_{{\rm pl}} = -\alpha\int {\rm d} V_{\rm n} \int {\rm d} V_{\rm e}
\left(
   \rho_{\rm n}(\vec r_{\rm n}) \, \rho_{\rm e}(\vec r_{\rm e})
   -\vec j_{\rm n}(\vec r_{\rm n}) \cdot \vec j_{\rm e}(\vec r_{\rm e})
   \right) \, \frac{e^{\iunit\omega| \vec r_{\rm n} - \vec r_{\rm e} |}}
   {\left| \vec r_{\rm n} - \vec r_{\rm e}\right|}
\quad,
\label{u}
\end{equation}
$\vec r_{\rm e}$ being the electronic coordinate, $\vec r_{\rm n}$ the
nuclear coordinate.

Since we neglect in our work the penetration of the electron wave functions
we do not have to specify the nuclear transition charge and current densities
$\rho_{\rm n}$ and $\vec j_{\rm n}$. The electronic transition charge and
current densities read
\begin{equation}
\rho_{\rm e} = \psi_{\rm f}^\dagger \, \psi_{\rm i}
\quad , \qquad
\vec j_{\rm e} = \psi_{\rm f}^\dagger \,\vec\alpha\, \psi_{\rm i}
\end{equation}
($\vec\alpha$ is the 3-vector of the spatial Dirac matrices in the standard
representation). These expressions are evaluated utilizing the scattering
solutions, see Eqs.~(\ref{eswelle},\ref{pswelle}) in the appendix, 
for the electron and positron wave 
function in order to define the emission direction and thus an opening angle.
Inserting the spherical wave expansion of
these wave functions results in a decomposition of the matrix element,
Eq.~(\ref{u}),
\begin{eqnarray}
   U_{\rm pl} = \sum\limits_{\kappa' , \mu'} \sum\limits_{\kappa , \mu}
   a^{(-) *}_{\kappa' \mu'} \; b^{(+)}_{\kappa \mu} \; 
   U_{\kappa'\mu' \kappa\mu}
   \quad .
\end{eqnarray}
$U_{ \kappa'\mu' \kappa\mu }$ 
denotes the transition matrix element which has the same structure
as $U_{\rm pl}$, but is evaluated
using the spherical spinor solutions of the Dirac equation, Eqs.~(\ref{ewelle}).
This matrix element was calculated in \cite{schlueter:81}. Here we cite the
result:
\begin{eqnarray}
   U_{ \kappa'\mu' \kappa\mu }
& = &
   4\pi\iunit\alpha\omega \, (-1)^{J_f-M_f}
   \left(\begin{array}{ccc} J_f & L & J_i \\ -M_f & M & M_i 
   \end{array}\right)
   V_{\gamma}^{(\tau)}(L)
\nonumber
\\
& &
   \times (-1)^{j'-\mu'}
   \left(\begin{array}{ccc} j' & L & j \\ -\mu' & M & \mu 
   \end{array}\right)
   M_{\kappa' \kappa}^{(\tau)}(L)
   \quad .
   \label{usph}
\end{eqnarray}
$V_{\gamma}^{(\tau)}(L)$ is just the reduced nuclear matrix element
of Eq.~(\ref{photon}) and
\begin{eqnarray}
M^{(\tau)}_{\kappa' \kappa}(L) 
& = &
-\iunit \, (-1)^{j'+\frac{1}{2}} \,
\frac{\sqrt{2j+1}\,\sqrt{2j'+1}\,\sqrt{2L+1}}{4\pi\sqrt{L(L+1)}}
\nonumber\\
& &
\times \left(\begin{array}{ccc} j & j' & L \\ -1/2 & 1/2 & 0 
\end{array}\right) 
R^{(\tau)}_{\kappa' \kappa}
\label{ematrix}
\end{eqnarray}
with the parity selection rule
\begin{equation}
l + l' + L + \lambda(\tau) = 0 \ \mbox{mod}\ 2 \quad, 
\quad
\left\{ \begin{array}{ccl} \lambda=0 & \mbox{for} & \tau = {\rm el} \\
                           \lambda=1 & \mbox{for} & \tau = {\rm magn}
\end{array}\right.
\end{equation}
$R^{(\tau)}_{\kappa' \kappa}$ contains the integration over the radial
electron wave functions and will be defined later.

Inserting this matrix element into the pair conversion probability, 
Eq.~({\ref p}), yields
\begin{eqnarray}
   P_{e^{+}e^{-}}
& = &
   2\pi
   \sum\limits_{M_i , M_i' , M_f} \rho^{[J_i]}_{M_i M_i'}
   \sum\limits_{\lambda , \lambda'}
   \int{\rm d} W\,{\rm d}\Omega \, \int{\rm d} W'\,{\rm d}\Omega \,
   \delta(\omega - W' - W)
\nonumber
\\
& &
   \times
   \sum\limits_{\kappa' , \mu'} \sum\limits_{\kappa , \mu}
   \sum\limits_{\bar\kappa' , \bar\mu'} \sum\limits_{\bar\kappa , \bar\mu}
   A_{\kappa' \mu' ; \bar\kappa' \bar\mu'} \, 
   B_{\kappa  \mu  ; \bar\kappa  \bar\mu } \, 
   U_{\kappa' \mu' \kappa \mu } \, 
   U^{*}_{\bar\kappa' \bar\mu' \bar\kappa \bar\mu }
   \quad .
\label{Pep}
\end{eqnarray}
where we abbreviated
\begin{equation}
   A_{\kappa' \mu' ; \bar\kappa' \bar\mu'} = W'\,p'\sum\limits_{\lambda'}
   a^{(-) *}_{\kappa' \mu'} \, a^{(-)}_{\bar\kappa' \bar\mu'}
\label{A}
\end{equation}
and
\begin{equation}
   B_{ \kappa  \mu  ; \bar\kappa  \bar\mu } = W\,p\sum\limits_{\lambda}
   b^{(+)}_{\kappa \mu} \, b^{(+) *}_{\bar\kappa \bar\mu}
   \quad .
\label{B}
\end{equation}
From Eq.~(\ref{Pep}) we obtain the differential pair conversion probability
with respect to the {\em kinetic positron energy} $E = W - m$ and the solid 
angles of both electron, $\Omega'$, and positron, $\Omega$,
\begin{equation}
P_{e^+e^-} = \int\limits_0^{\omega-2m} {\rm d}E \int{\rm d}\Omega
\int{\rm d}\Omega' \, \frac{{\rm d}^3 P_{e^+ e^-}}{{\rm d}E \,
{\rm d}\Omega \, {\rm d}\Omega'}
\quad .
\label{int}
\end{equation}
The integration over the electron energy $W'$ is trivially performed because 
of the $\delta$ function. From this relation it is obvious that we may proceed
from the solid angles $\Omega$ to $\tilde\Omega$ by choosing another 
reference axis in space. The integrand in Eq.~(\ref{int}) is invariant 
under rotations since the Jacobian of this transformation equals 1. The
integrand should thus be represented by a series of triple correlation functions
which are defined in Eq.~(\ref{tripelkorrelationsfunktion}). 

Inserting the explicit expressions of the coefficients $A$ and $B$,
Eqs.~(\ref{koeff-a},\ref{koeff-b}) leads to the following expression 
for the differential pair conversion probability
\begin{eqnarray}
\lefteqn{
\frac{{\rm d}^3 P_{e^+ e^-}}{{\rm d}E\,{\rm d}\Omega\,{\rm d}\Omega'}
=
8(\pi\alpha\omega)^2 \, |V_\gamma^{(\tau)}(L)|^2
\sum\limits_{M , \bar M} \sum\limits_{M_i , M_i'} \sum\limits_{M_f}
\rho^{[J_i]}_{M_i M_i'}
}
\nonumber
\\
& &
   \times
   \left(\begin{array}{ccc} J_f & L & J_i \\ -M_f & M & M_i 
   \end{array}\right)
   \left(\begin{array}{ccc} J_f & L & J_i \\ -M_f & \bar M & M_i' 
   \end{array}\right)
\nonumber
\\
& &
   \times
   \sum\limits_{\kappa , \kappa' , \bar\kappa , \bar\kappa'}
   (-1)^{j'+\bar j'} \, M^{(\tau)}_{\kappa' \kappa}(L) \,
   M^{(\tau) *}_{\bar\kappa' \bar\kappa}(L)
   \sqrt{2\bar j'+1} \, \sqrt{2j'+1}
\nonumber
\\
& &
   \times
   \sqrt{2j+1} \, \sqrt{2\bar j+1} \,
\nonumber 
\\
& &
   \times
   \exp(\iunit[(\delta'(W',\kappa')-\bar\delta'(W',\bar\kappa')
             +\delta(-W,\kappa)-\bar\delta(-W,\bar\kappa)])
\nonumber
\\
& &
   \times \sum\limits_{I' , I , \alpha , \alpha'}
   \sqrt{2I'+1} \, \sqrt{2I+1} \,
   Y_{I'\alpha'}(\Omega_{p'}) \, Y_{I\alpha}(\Omega_{p})
\nonumber
\\
& &
   \times
   \left(\begin{array}{ccc} j' & \bar j' & I' \\ 1/2 & -1/2 & 0 
   \end{array}\right)
   \left(\begin{array}{ccc} j & \bar j & I \\ 1/2 & -1/2 & 0 
   \end{array}\right)
\nonumber
\\
& &
   \times
   \sum\limits_{\mu , \mu' , \bar\mu , \bar\mu'}
   (-1)^{\bar\mu-\bar\mu'+1}
   \left(\begin{array}{ccc} \bar j' & L & \bar j \\ -\bar\mu' & \bar M & \bar\mu 
   \end{array}\right)
\nonumber
\\
& &
   \times
   \left(\begin{array}{ccc} j' & L & j \\ -\mu' & M & \mu 
   \end{array}\right)
   \left(\begin{array}{ccc} \bar j' & j' & I' \\ -\bar\mu' & \mu' & -\alpha' 
   \end{array}\right)
   \left(\begin{array}{ccc} \bar j & j & I \\ -\bar\mu & \mu & \alpha 
   \end{array}\right)
   \quad .
\end{eqnarray}
Here we inserted Eq.~(\ref{usph}).
Introducing the statistical tensors we are left with
\begin{eqnarray}
   \frac{{\rm d}^3 P_{e^+ e^-}}{{\rm d}E\,{\rm d}\Omega\,{\rm d}\Omega'}
& = &
   2(4\pi\alpha\omega)^2 \, |V_\gamma^{(\tau)}(L)|^2 \,
   (-1)^{J_f-J_i+L+1} 
\nonumber
\\
& &
   \times
   \sum\limits_{n , \nu} \sqrt{2n+1} (-1)^{\nu} \hat\rho^{[n]}_{\nu}(J_i)
   \left\{\begin{array}{ccc} L & L & n \\ J_i & J_i & J_f \end{array}\right\}
\nonumber
\\
& &
   \times \sum\limits_{I , I'} 
   \sqrt{2I+1} \, \sqrt{2I'+1} \, (-1)^{I'} \,
\nonumber
\\
& &
   \times
   \sum\limits_{\alpha , \alpha'}
   Y_{I'\alpha'}(\Omega_{p'}) \, Y_{I\alpha}(\Omega_{p})
   \left(\begin{array}{ccc} I & I' & n \\ \alpha & \alpha' & -\nu 
   \end{array}\right)
\nonumber
\\
& &
   \times
   \sum\limits_{\kappa , \kappa' , \bar\kappa , \bar\kappa'}
   (-1)^{\bar j+\bar j'} \, 
   \sqrt{|\kappa \, \kappa' \, \bar\kappa \, \bar\kappa'|}
   \left\{\begin{array}{ccc} \bar j & \bar j' & L \\
               j & j' & L \\
               I & I' & n \end{array}\right\}
   M^{(\tau)}_{\kappa' \kappa}(L) \, M^{(\tau) *}_{\bar\kappa' \bar\kappa}(L)
\nonumber
\\
& &
   \times
   \exp(\iunit[\delta'(E',\kappa')-\bar\delta'(E',\bar\kappa')+
             \delta(-E,\kappa)-\bar\delta(-E,\bar\kappa)])
\nonumber
\\
& &
   \times
   \left(\begin{array}{ccc} j' & \bar j' & I' \\ 1/2 & -1/2 & 0 
   \end{array}\right)
   \left(\begin{array}{ccc} j & \bar j & I \\ 1/2 & -1/2 & 0 
   \end{array}\right)
   \quad.
\end{eqnarray}
This is the most general form for the pair conversion probability. Now
we assume that we are dealing with internal pair conversion of aligned
nuclei ($\nu=0$). 
We may choose an appropriate coordinate system by transformation
of the spherical harmonics:
\begin{eqnarray}
   Y_{I\alpha}(\Omega_{p}) 
& = & 
   \sum\limits_{\beta}
   \exp(\iunit\alpha\phi) \, d^{[I]}_{\alpha\beta}(\vartheta) \,
   \exp(\iunit\beta\delta) \, Y_{I\beta}( 0, 0 )
\nonumber\\
& = &
   \sqrt{\frac{2I+1}{4\pi}} \, \exp(\iunit\alpha\phi) \,
   d^{[I]}_{\alpha 0}(\vartheta)
\nonumber
\\
   Y_{I'\,-\alpha}(\Omega_{p'})
& = &
   \sum\limits_{\beta'} \exp(-\iunit\alpha\phi) \,
   d^{[I']}_{-\alpha\beta'}(\vartheta) \, \exp(\iunit\beta'\delta)
   \, Y_{I'\beta'}(\Theta, 0)
\end{eqnarray}

Here $\Theta$ denotes the opening angle of the electron-positron pair,
$\vartheta$ is the polar angle of the positron with respect to the symmetry
axis, and the dihedral angle $\delta$ indicates the rotation of the 
electron-positron plane around the positron axis (Fig.~2b). Please note,
that the convention of \cite{goldring,rose:63,warburton} differs in the
definition of the angles from the one employed here.

This enables us to define the triple correlation function by
\begin{eqnarray}
\lefteqn{P_{II'n}(\vartheta,\Theta,\delta) = 
   \sum\limits_{\alpha}
   \left(\begin{array}{ccc} I & I' & n \\ \alpha & -\alpha & 0 
   \end{array}\right)
   Y_{I\alpha}(\Omega_{p}) \, Y_{I'\,-\alpha'}(\Omega_{p'})
}
\nonumber
\\
& = &
   \frac{\sqrt{2I+1} \, \sqrt{2I'+1}}{4\pi}
   \sum\limits_{\beta'} (-1)^{\beta'}
   \left(\begin{array}{ccc} I & I' & n \\ 0 & \beta & \beta' 
   \end{array}\right)
   d^{[n]}_{\beta' 0}(\vartheta) \, d^{[I']}_{\beta' 0}(\Theta) \,
   \exp(\iunit\beta'\delta)
   \quad .
\nonumber
\\
& &
\label{tripelkorrelationsfunktion}
\end{eqnarray}
Our triple correlation function is related to the one introduced
by Biedenharn \cite{biedenharn} by a factor
$4\pi\iunit^{-I-I'-n}/((2I+1)(2I'+1))^{1/2}$.

The pair conversion probability is normalized by the probability for
$\gamma$ emission, Eq.~(\ref{photon}), which yields the pair conversion
coefficient:
\begin{eqnarray}
\lefteqn{
   \frac{{\rm d}^4 \beta}{{\rm d}E\,{\rm d}\cos\Theta\,{\rm d}\cos\vartheta
   \,{\rm d}\delta} =
   \frac{2\alpha\omega (2L+1)}{L(L+1)} \, (2J_i+1) \,
   (-1)^{J_f-J_i+L+1} } 
\nonumber
\\
& &
   \times
   \sum\limits_{n} \sqrt{2n+1} \, \hat\rho^{[n]}_{0}(J_i) \,
   \left\{\begin{array}{ccc} L & L & n \\ J_i & J_i & J_f 
   \end{array}\right\}
\nonumber
\\
& &
   \times \sum\limits_{I , I'} (2I+1) (2I'+1) (-1)^{I'}
\nonumber
\\
& &
   \times \sum\limits_{\beta'} (-1)^{\beta'}
   \left(\begin{array}{ccc} I & I' & n \\ 0 & \beta' & -\beta' 
   \end{array}\right)
   d^{[n]}_{\beta' 0}(\vartheta) \, d^{[I']}{\beta' 0}(\Theta) \,
   \exp(\iunit\beta'\delta)
\nonumber
\\
& &
   \times \sum\limits_{\kappa , \kappa' , \bar\kappa , \bar\kappa'}
   (-1)^{j+\bar j'} |\kappa \, \kappa' \, \bar\kappa \, \bar\kappa'|
   \left\{\begin{array}{ccc} \bar j & \bar j' & L \\ j & j' & L \\ 
   I & I' & n \end{array}\right\}
   R^{(\tau)}_{\kappa' \kappa}(L) \, R^{(\tau) *}_{\bar\kappa'\bar\kappa}(L)
\nonumber
\\
& &
   \times \exp(\iunit[\delta'(W',\kappa')-\bar\delta'(W',\bar\kappa')+
                    \delta(-W,\kappa)-\bar\delta(-W,\bar\kappa)]) \,
\nonumber
\\
& &
   \times
   \left(\begin{array}{ccc} j' & \bar j' & I' \\ 1/2 & -1/2 & 0 
   \end{array}\right)
   \left(\begin{array}{ccc} j & \bar j & I \\ 1/2 & -1/2 & 0 
   \end{array}\right)
\nonumber
\\
& &
   \times
   \left(\begin{array}{ccc} j & j' & L \\ 1/2 & -1/2 & 0 
   \end{array}\right)
   \left(\begin{array}{ccc} \bar j & \bar j' & L \\ 1/2 & -1/2 & 0 
   \end{array}\right)
   \quad .
   \label{ergebnis}
\end{eqnarray}
Here we inserted the explicit expressions for the electronic matrix
elements, Eq.~(\ref{ematrix}). Integration over the azimuthal angle is
trivially performed resulting in an additional factor of $2\pi$.

The radial matrix elements read for electric pair conversion (parity
$(-)^{L}$)
\begin{equation}
R^{({\rm e})}_{\kappa' \kappa} =
L ( R_1 + R_2 + R_3 - R_4 ) + ( \kappa - \kappa' ) ( R_3 + R_4 )
\label{erade}
\end{equation}
and for magnetic pair conversion (parity $(-)^{L+1}$)
\begin{equation}
R^{({\rm m})}_{\kappa' \kappa} =
(\kappa + \kappa') ( R_5 + R_6 )
\quad .
\label{eradm}
\end{equation}
The radial integrals introduced in these equations are taken over products of
the radial electron wave functions (\ref{radiale}) and the Hankel functions
of first kind, $h^{(1)}_L(\omega r)$:
\begin{eqnarray}
R_1 & = & \int\limits_0^\infty {\rm d} r \, r^2 \, g_{W',\kappa'}(r) \, 
g_{-W,\kappa}(r)
\,
          h^{(1)}_L(\omega r) \quad , \nonumber\\
R_2 & = & \int\limits_0^\infty {\rm d} r \, r^2 \, f_{W',\kappa'}(r) \, 
f_{-W,\kappa}(r)
\,
          h^{(1)}_L(\omega r) \quad , \nonumber\\
R_3 & = & \int\limits_0^\infty {\rm d} r \, r^2 \, g_{W',\kappa'}(r) \, 
f_{-W,\kappa}(r)
\,
          h^{(1)}_{L-1}(\omega r) \quad , \nonumber\\
R_4 & = & \int\limits_0^\infty {\rm d} r \, r^2 \, f_{W',\kappa'}(r) \, 
g_{-W,\kappa}(r)
\,
          h^{(1)}_{L-1}(\omega r) \quad , \nonumber\\
R_5 & = & \int\limits_0^\infty {\rm d} r \, r^2 \, g_{W',\kappa'}(r) \, 
f_{-W,\kappa}(r)
\,
          h^{(1)}_L(\omega r) \quad , \nonumber\\
R_6 & = & \int\limits_0^\infty {\rm d} r \, r^2 \, f_{W',\kappa'}(r) \, 
g_{-W,\kappa}(r)
\,
          h^{(1)}_L(\omega r)
\label{radialintegrale}
\quad.
\end{eqnarray}
In the case of a point-like nucleus these integrals can be rewritten in
terms of $F_2$ functions \cite{schlueter:78} which can be evaluated numerically.
For the representation of the nucleus as a homogeneously charged
sphere the radial integrals are computed using a Gauss-Chebyshev quadrature
\cite{chebyshev}. 
The Whittaker functions which occur in the expressions for the electron
wave functions are computed with the COULCC code of \cite{barnett}.
Since the integrands are oscillating functions 
it is advantageous to deform the
integration contour in the complex plane 
in such a way that it runs along the imaginary axis
\cite{schlueter:81}.
Since the electron wave functions have the asymptotic behaviour 
$\exp(\iunit p r)$ while the Hankel functions behave like 
$\exp(\iunit \omega r)$, where $\omega=W+W'$, the integrand for
large $r$ assumes the form: 
\begin{equation}
\exp(\iunit[-p-p'+W+W']r)
\quad .
\end{equation}
For $r$ complex with the imaginary part going to infinity, our procedure thus
guarantees that the integrand falls off quite fast. In most cases
at $r=20000$ fm the integrand is smaller than $10^{-5}$ of its maximum value.

We want to consider the angular correlation for two special cases: 

I.
If we integrate Eq.~(\ref{ergebnis}) 
over the positron polar angle $\vartheta$ and the dihedral
angle $\delta$ the remaining function depends only on the opening angle
$\Theta$ of the electron-positron pair. In this case 
only the $n=0$ contribution survives. 
We get the opening angle
distribution as a series of Legendre polynomials which was already calculated 
in \cite{hofmann:90}:
\begin{equation}
\frac{{\rm d}^2\beta}{{\rm d}E\,{\rm d}\cos\Theta} = 
\sum\limits_{I} a_I \, P_I(\cos\Theta)
\end{equation}
The expansion coefficients are given by
\begin{eqnarray}
a_I & = &
\frac{8\pi\alpha\omega}{L(L+1)} \, (-)^{L+I+1} (2I+1) 
\sum\limits_{\kappa,\kappa',\bar\kappa,\bar\kappa'} \left|
\kappa\,\kappa'\,\bar\kappa\,\bar\kappa'\right| 
R^{(\tau)}_{\kappa'\kappa}(L) \, R^{(\tau)*}_{\bar\kappa'\bar\kappa}
\nonumber\\
& \times &
\exp(\iunit[\delta'(W',\kappa') - \bar\delta'(W',\bar\kappa')
               +\delta(-W,\kappa) - \bar\delta(-W,\bar\kappa)])
\nonumber\\
& \times &
\left(\begin{array}{ccc} j' & \bar j' & I \\ 1/2 & -1/2 & 0 \end{array}\right)
\left(\begin{array}{ccc} j & \bar j & I \\ 1/2 & -1/2 & 0 \end{array}\right)
\nonumber\\
& \times &
\left(\begin{array}{ccc} j & j' & L \\ 1/2 & -1/2 & 0 \end{array}\right)
\left(\begin{array}{ccc} \bar j & \bar j' & L \\ 1/2 & -1/2 & 0
\end{array}\right)
\left\{\begin{array}{ccc} \bar j & \bar j' & L \\ j' & j & L \end{array}
\right\}
\quad .
\end{eqnarray}
They have to be evaluated numerically. 
The same result is achieved if one assumes that the initial nuclear
substates are equally populated.

At this point we apologize for
giving an incorrect expression for the opening angle distribution in
\cite{hofmann:90} which was caused by employing the wrong set of
scattering solutions. 
This error resulted in the wrong sign
of the scattering phase shifts of the positron. The opening angle
distribution showed the right qualitative behaviour but wrong conversion
probabilities. The statement, that the maximum of the distribution shifts
from $0^\circ$ to $180^\circ$, if one considers
overcritical nuclear charges ($Z\geq 173$), remains unchanged. This
error appeared also in the expression for the electric monopole (E0)
conversion. One should reverse the sign of the scattering phase shifts
of the positron. The pair conversion coefficient for
the electric monopole conversion reads:
\begin{equation}
\frac{{\rm d}\eta}{{\rm d}E \, {\rm d}\cos\Theta} 
= \frac{1}{2} \, \frac{{\rm d}\eta}{{\rm d}E}
( 1 + \varepsilon \, \cos\Theta )
\label{maximum}
\end{equation}
where ${\rm d}\eta / {\rm d}E$ is the differential
pair conversion coefficient \cite{soff:81} 
---which remains unchanged--- and $\varepsilon$ is the
corrected anisotropy coefficient:
\begin{equation}
\varepsilon = 2 \, \frac{C_{-1}\,C_{+1}}{C_{-1}^2 + C_{+1}^2} \,
\cos(\Delta_{+1 \, -1})
\end{equation}
with
\begin{equation}
\Delta_{\kappa\kappa'} = 
\delta'(W',\kappa)-\delta'(W',\kappa')+
\delta(-W,\kappa)-\delta(-W,\kappa')
\quad.
\end{equation}

$C_{+1}$, $C_{-1}$ are defined by
\begin{equation}
C_\kappa = \left\{ \begin{array}{ccc}
\displaystyle\lim_{r\to 0} \frac{f_{-W,\kappa}(r) \, f'_{W',\kappa}(r)}
{r^{2j-1}} & \mbox{for} & \kappa>0
\\[3mm]
\displaystyle\lim_{r\to 0} \frac{g_{-W,\kappa}(r) \, f'_{W',\kappa}(r)}
{r^{2j-1}} & \mbox{for} & \kappa<0
\end{array}\right .
\end{equation} 
where $f$ ($f'$) and $g$ ($g'$) are the radial wave functions of the 
Dirac spinor of the positron (electron). In numerical calculations these
constants are evaluated at the nuclear radius.
$\delta$ and $\delta'$ are the corresponding Coulomb phase shifts
for an extended nucleus \cite{mueller:rafelski:greiner:3}.

II. 
If we integrate Eq.~(\ref{ergebnis}) over the opening angle $\Theta$ of the electron-positron 
pair and the dihedral angle $\delta$ we end up with
\begin{equation}
\frac{{\rm d}^2\beta}{{\rm d}E\,{\rm d}\cos\vartheta} =
\sum\limits_n b_n \, P_n(\cos\vartheta)
\end{equation}
where the coefficients read
\begin{eqnarray}
b_n & = &
4\pi\alpha\omega \, \frac{(2L+1)(2J_i+1)}{L(L+1)} \, (-)^{J_f - J_i + 1}
\sqrt{2n+1} \, \hat\rho^{[n]}_0(J_i)
\left\{\begin{array}{ccc} L & L & n \\ J_i & J_i & J_f \end{array}\right\}
\nonumber\\
& \times &
\sum\limits_{\kappa,\kappa',\bar\kappa} (-)^{j+\bar j + j'+1/2}
\left|\kappa\,\kappa'\,\bar\kappa\right|
\left\{\begin{array}{ccc} L & L & n \\ j & \bar j & j' \end{array}\right\}
R^{(\tau)}_{\kappa'\kappa}(L) \, R^{\tau *}_{\kappa'\bar\kappa}(L)
\nonumber\\
& \times &
\exp(\iunit[\delta(-W,\kappa) - \bar\delta(-W,\bar\kappa)])
\nonumber\\
& \times &
\left(\begin{array}{ccc} j & \bar j & n \\ 1/2 & -1/2 & 0 \end{array}\right)
\left(\begin{array}{ccc} j & j' & L \\ 1/2 & -1/2 & 0 \end{array}\right)
\left(\begin{array}{ccc} \bar j & j' & L \\ 1/2 & -1/2 & 0 \end{array}\right)
\quad .
\end{eqnarray}
This corresponds to the experimental setup where one is just interested
in the angular distribution of the positron emitted in internal pair
conversion of an aligned or polarized nucleus.
%
%
\section{Results}
%
%
In the following we will discuss the characteristic properties of IPC
angular distributions using a few representative results. Since we
are interested in Coulomb effects they all refer to a uranium-like 
nucleus ($Z=92$). The chosen energies and multipolarities are
generic and are not intended to represent particular nuclear
transitions known from experiment.

The opening angle distribution of electron and positron emitted by 
IPC depicts for electric transitions the typical pattern: 
it has its maximum at $\Theta=0^\circ$ and
its minimum for $\Theta=180^\circ$. For magnetic transitions in
heavy nuclei, however, the situation might be different.
Fig.~1a and 1b depict the
opening angle distribution for an E1 and a M1 transition of a uranium-like
nucleus as a result of our distorted wave Born approximation (DWBA)
in comparison with the Born approximation (BA). This demonstrates
how the angular correlation of the electron-positron pairs is
influenced by the strong Coulomb field of the nucleus.
In Fig.~1b we plotted also the opening-angle distribution
for the M1 transition taking into account the finite extension of the
nucleus under consideration. This verifies that the magnetic transitions
---and especially the M1 transition--- are very sensitive to the charge
distribution of the nucleus \cite{schlueter:81}. 
For the E1 transition in Fig.~1a, on the other hand, the effect of the finite 
nuclear size amounts to less than 0.1\,\%.

In the following we discuss
the triple angular correlation of electron and positron
for IPC of aligned nuclei. We take the symmetry axis as quantization axis
as in Eq.~(\ref{ergebnis}). The opening angle $\Theta$ of electron
and positron, the polar angle $\vartheta$ of the positron and the dihedral
angle $\delta$ form a complete set of angles to fix the emission directions
of electron and positron with respect to the symmetry axis. 
The angles which describe the directions of the emitted leptons are displayed
in Fig.~2b. Note that our
choice of the coordinate system is different from that introduced 
in the Born approximation calculations of
\cite{goldring,rose:63,warburton} in which $\vartheta$ denotes the polar
angle of the intermediate photon. However, since the Coulomb field 
disturbs the momentum balance
we cannot determine the momentum of the intermediate
photon from the momenta of the outgoing leptons, which 
would be necessary to
calculate the photon polar angle.

Depending on the experimental setup and reactions 
various coordinate systems may be established in which the 
statistical tensors are determined. Here we concentrate on
the Coulomb excitation of heavy ions in collisions 
with beam energies at or below the
Coulomb barrier. In this case one usually chooses a coordinate system,
where the $z$ axis is pointing along the apex line 
of the scattering hyperbola towards the projectile
and the $x$ axis is perpendicular to the scattering plane
(Fig.~2a). The $y$ axis
is then chosen such that the $y$ component of the projectile velocity
is positive \cite{alder:winther:1,alder:winther:2,broglia}. 
In the sudden approximation it can be shown 
that the nuclear states are excited with 
a population of the magnetic substates reaching a maximum around
$M_i=0$, i.e.,
the nucleus is aligned in the plane perpendicular to the $z$ axis
(asymptotic recoil direction of the target)
\cite{wollersheim}. This is called oblate
alignment. Taking into account the de-excitation of the nucleus
by $\gamma$ cascades starting from high-spin the oblate alignment
changes into a prolate alignment with respect to the $z$ axis for the
low-spin states.

If the collision energy is increased the nuclear alignment changes
to a polarization with respect to a reference axis perpendicular to the
scattering plane \cite{alder:winther:2,broglia}. Classically
this corresponds to the situation where the drag caused by 
surface friction puts the nuclei into a spinning motion.

After having chosen a coordinate system and having determined the degree
of alignment or polarization for the Coulomb excited nuclei 
---the corresponding statistical tensors can be calculated with, e.g.,
the COULEX code of \cite{alder:winther:1}--- one can
employ Eq.~(\ref{ergebnis}) to determine the angular distribution of the
electron-positron pairs emitted by internal pair conversion of these
nuclei. 
We plot in Fig.~3 the spatial correlation of the 
electron-positron
pairs with respect to the reference axis assuming oblate alignment
of a uranium-like
nucleus. From the spectrum of the emitted pairs \cite{schlueter:81,schlueter:79}
we know that for large-$Z$ nuclei the
pair emission probability increases towards the maximum positron energy. Thus
the angular correlations are plotted for a case where nearly 
the full transition energy (minus the electron rest mass)
is transferred to the positron. One recognizes a strong dependence
of the pair conversion probability on the polar angle of the positron with
respect to the reference axis. This behaviour resembles the anisotropic
emission of the intermediary photon
\cite{wollersheim}. The angular distribution
depends weakly on the dihedral angle $\delta$ of the electron-positron pair
(Fig.~4).
For transitions between nuclear states of high angular momentum the
opening angle distribution does not change drastically when the positron
polar angle is varied. 

In order to elucidate the influence of the statistical tensors,
i.e., the occupation of the initial nuclear state on
the angular correlation of the emitted electron-positron pairs,
we present the angular distribution with respect to the polar
angle of the emitted positron.
Fig.~5 shows the polar angle
distribution assuming E1, E2 and E3 transitions to the $0^+$
ground state of nuclei which exhibit oblate
alignment. Fig.~6 displays the polar angle distribution
for a E1 transition to the $0^+$ ground state of a nucleus 
for oblate and prolate alignment and for
polarization. 
\section{Conclusion}
Angular correlations are very sensitive to the underlying process.
They may reveal a plethora of signatures for nuclear transition which
allow for an identification of the transition as well as for the study of the
properties of excited nuclei. Especially for large-$Z$ nuclei one cannot
rely on the validity of the Born approximation which becomes exact in
the limit $Z\rightarrow 0$. One rather has to perform the calculations
with the relativistic scattering wave functions for both electron and positron.
These wave functions take the 
Coulomb distortion caused by the nuclear charge into account.

For magnetic transitions of large-$Z$ nuclei one has also to account for
the finite-size effects. Especially M1 transitions are very
sensitive on the extension of the nucleus. In order to study magnetic
IPC we approximated the nucleus undergoing the transition
by a homogeneously charged sphere.
The angular correlation of electron-positron pairs with respect to a
given axis in space depend on the statistical tensors which reflect the
population of the nuclear magnetic substates. The conversion probability
changes drastically when either the opening angle of the pair or the
polar angle of the positron is varied while the dependence on the
dihedral angle is rather weak. 
Our calculations allow to make quantitative predictions of this 
behaviour which qualitatively might have been anticipated from
the $\gamma$-ray spectroscopy performed in heavy-ion collisions. 
Furthermore, the measurement
of the spatial correlation of electron-positron pairs can be 
employed to obtain additional information about the nuclear transition.
E.g., not only the multipolarity but also the parity of the nuclear
transition can be measured in this way. 
\\[1cm]

{\bf Acknowledgement:}
This work has been supported by the BMBF, by the 
Deut\-sche For\-schungs\-gemein\-schaft (DFG), by GSI (Darmstadt),
and by the REHE programme of the European Science Foundation (ESF).
\newpage
\appendix
%
%
\section{Electron wave functions for point-like and\protect\newline 
         extended nuclei}
%
%

In our calculations we employed the following form of the spherical continuum
wave functions of the electron moving in the Coulomb field of a point-like
nucleus \cite{rose:61}:
\begin{equation}
   \chi_{W, \kappa, \mu}(\vec r\,)=\left(\begin{array}{c}
   g_{W,\kappa}(r)\,\chi_{\kappa\mu}(\hat r) \\
   \iunit f_{W,\kappa}(r)\,\chi_{-\kappa\mu}(\hat r)
   \end{array}\right) \quad.
   \label{ewelle}
\end{equation}
The spinor spherical harmonics are defined as
\begin{equation}
\chi_{\kappa\mu}(\Omega) =
\sum\limits_{\kappa , \mu}
\left(
\begin{array}{cc|c} l & \halb & j \\ m & \lambda & \mu \end{array}\right)
Y_{\lambda m}(\Omega) \, \chi_{\lambda}
\end{equation}
where the basis spinors are given as usual by
\begin{equation}
\chi_{\halb} = \left( \begin{array}{c} 1 \\ 0 \end{array} \right) \quad , 
\qquad
\chi_{-\halb} = \left( \begin{array}{c} 0 \\ 1 \end{array} \right) \quad.
\end{equation}
Defining the relativistic Sommerfeld parameter as $y=-Z\alpha W/p$ where
$p=\sqrt{W^2-m^2}$, the radial wave functions read
for a point nucleus
\begin{eqnarray}
   g_{W,\kappa}(r)
   & = &
   \sqrt{\frac{W+m}{\pi p}} \frac{1}{r}
   \frac{|\Gamma(\gamma-\iunit y)|}
   {2\Gamma(2\gamma+1)}\,e^{-\pi y/2}\,(2pr)^\gamma\nonumber
   \\
   & &
   \times\left\{(\gamma-\iunit y)\,e^{-\iunit(pr-\eta)}\,
   _1F_1(\gamma+1-\iunit y, 2\gamma+1; 2\iunit pr) + {\rm c.c.}\right\}
   \quad,
   \nonumber\\
   f_{W,\kappa}(r)
   & = &
   \sqrt{\frac{W-m}{\pi p}} \frac{\iunit}{r}
   \frac{|\Gamma(\gamma-\iunit  y)|}
   {2\Gamma(2\gamma+1)}\,e^{-\pi y/2}\,(2pr)^\gamma \nonumber\\
   & &
   \times\left\{(\gamma-\iunit y)\,e^{-\iunit(pr-\eta)}\,
   _1F_1(\gamma+1-\iunit y, 2\gamma+1; 2\iunit pr) - {\rm c.c.}\right\}
   \quad.
\label{radiale}
\end{eqnarray}
with
$$
   \eta = \onehalf\arg\left(-\frac{\kappa+\iunit ym/W}{\gamma-\iunit y}\right)
   \quad,\qquad
   \gamma=\sqrt{\kappa^2 - (Z\alpha)^2} \quad.
$$

For an extended nucleus we construct the continuum solutions
as in \cite{mueller:rafelski:greiner:3}
by employing a power series ansatz for the electron wave function inside the
nucleus which is matched to the linear combination of wave functions to the
Coulomb potential at the nuclear radius. From the matching condition the 
normalization factor and the phase shift can be deduced.


The continuum solutions of the Dirac equation can be written
as wave functions which asymptotically represent 
plane waves of momentum $\vec p$ and spin $\lambda$.
These wave functions are obtained as a series expansion into spherical
harmonics \cite{rose:61,hofmann:93}:

For positive energies this expansion reads: 
\begin{equation}
   \label{e30}
   \psi_{W,\vec p,\lambda}^{(\pm)}=\sum\limits_{\kappa,\mu}
   a^{(\pm)}_{\kappa\mu}\,\chi_{W,\kappa,\mu}
   \quad.
   \label{eswelle}
\end{equation}
with the coefficients
\begin{equation}
   a^{(\pm)}_{\kappa\mu} = 
   \frac{1}{\sqrt{Wp}}\,\iunit^l
   e^{\pm\iunit[\delta(W,\kappa) + \pi(l+1)/2]}
   \sum\limits_m Y^*_{lm}(\hat p)
   \left(\begin{array}{cc} l & \onehalf \\ m & \lambda \end{array}
   \left| \begin{array}{c} j \\ \mu \end{array} \right.\right)
   \label{esa}
\end{equation}
and the Coulomb phase shift 
\begin{equation}
\delta(W,\kappa) 
= \eta - \arg\Gamma(\gamma - \iunit y) - \frac{\pi}{2} \, \gamma
\end{equation}
and for negative energies ($-W<0$) \cite{hofmann:93}
\begin{equation}
   \psi^{(\pm)}_{-W,\vec p, \lambda} = \sum\limits_{\kappa,\mu}
   b^{(\pm)}_{\kappa\mu}\,\chi_{-W,\kappa,\mu}
   \label{pswelle}
\end{equation}
with the coefficients
\begin{equation}
   b^{(\pm)}_{\kappa, \mu} = 
   \frac{1}{\sqrt{Wp}}\,\iunit^{(l(-\kappa) +1)}
   e^{\pm\iunit[\delta(-W,\kappa) +\pi l(-\kappa)]}
   \sum_{m} Y_{l(-\kappa) m}^* (\hat p)
   \left(\begin{array}{cc} l(-\kappa) & \onehalf \\ m & \lambda \end{array}
   \left| \begin{array}{c} j \\ \mu \end{array} \right.\right)
   \label{psb}
\end{equation}
and the Coulomb phase shift 
\begin{equation}
\delta(-W,\kappa) 
= \eta - \arg\Gamma(\gamma - \iunit y) - \frac{\pi}{2} \, \gamma
\quad .
\end{equation}
These wave functions obey the normalization condition
\begin{equation}
\int {\rm d}^3r\,\left(\psi^{(\pm)}_{W,\vec p, \lambda}(\vec r\,)
\right)^{\dag}\,\psi^{(\pm)}_{W',\vec p\,', \lambda'}(\vec r\,)
=\delta^3(\vec p-\vec p\,') \, \delta_{\lambda\,\lambda'}
\quad .
\end{equation}
In the case of an extended nucleus the phase shifts have to be determined
numerically.
%
%
\section{Evaluation of the coefficients $A$ and $B$}
%
%
We start with the coefficients $A$ which contain the electron scattering
wave function and its complex conjugate. Inserting the explicit form of these
coefficients, Eq.~(\ref{esa}), we get
\begin{eqnarray}
A_{ \kappa'\mu' ; \bar\kappa'\bar\mu' } 
& = &
\sum\limits_{\lambda' , m' , \bar m'} 
\exp(\iunit[\delta'(W',\kappa')-\bar\delta'(W',\bar\kappa')]) 
\nonumber
\\
& \times & \sqrt{2j'+1} \, \sqrt{2\bar j'+1} 
(-1)^{l'(\kappa')-1/2 +\mu'}
\nonumber
\\
& \times &
Y_{l'(\kappa')m'}(\Omega_{p'}) \, Y_{\bar l'(\bar\kappa')\bar m'}^*
(\Omega_{p'}) \, (-1)^{\bar l'(\bar\kappa')-1/2+\bar\mu'}
\nonumber
\\
& \times &
\left(\begin{array}{ccc} l'(\kappa') & 1/2 & j' \\ m' & \lambda' & -\mu' 
\end{array}\right)
\left(\begin{array}{ccc} \bar l'(\bar\kappa') & 1/2 & \bar j' \\ 
\bar m' & \lambda' & -\bar\mu' \end{array}\right)
\quad ,
\label{winkel-9}
\end{eqnarray}
which can be transformed into
\begin{eqnarray}
\lefteqn{
   A_{ \kappa'\mu' ; \bar\kappa'\bar\mu' }
   = \frac{1}{\sqrt{4\pi}} 
   \exp(\iunit[\delta'(W',\kappa')-\bar\delta'(W',\bar\kappa')]) 
}
\nonumber
\\
   & \times &
   (-1)^{\mu'+1/2} \, \sqrt{2j'+1} \, \sqrt{2\bar j'+1}
\nonumber
\\
   & \times &
   \sum\limits_{I' , \alpha'} \sqrt{2I'+1} \; Y_{I'\alpha'}(\Omega_{p'})
   \left(\begin{array}{ccc} j' & \bar j' & I' \\ 1/2 & -1/2 & 0 
   \end{array}\right)
   \left(\begin{array}{ccc} \bar j' & j' & I' \\ -\bar\mu' & \mu' & -\alpha' 
   \end{array}\right)
   \quad .
\label{koeff-a}
\end{eqnarray}
Additionally we get the parity selection rule
$$l'(\kappa') + \bar l'(\bar\kappa') + I' = 0 \mbox{ mod } 2 \quad,$$
and the angular momentum selection rule:
$$|l'(\kappa') - \bar l'(\bar\kappa')| \leq I' 
\leq l'(\kappa') + \bar l'(\bar\kappa') \quad.$$
Next we proceed to evaluate the coefficients $B$, which are composed from
the positron scattering wave functions, Eq.~(\ref{psb}):
\begin{eqnarray}
B_{ \kappa\mu ; \bar\kappa\bar\mu } 
& = &
\sum\limits_{m , \bar m , \lambda}
(-1)^{1+\mu+\bar\mu} \, 
\exp(\iunit[\delta(-W,\kappa)-\delta(-W,\bar\kappa)]) 
\nonumber
\\
& \times &
\sqrt{2j+1} \, \sqrt{2\bar j+1}
Y_{l(-\kappa)m}^*(-\Omega_{p}) \, 
Y_{\bar l(-\bar\kappa)\bar m}(-\Omega_{p})
\nonumber
\\
& \times &
\left(\begin{array}{ccc} l(-\kappa) & 1/2 & j \\ 
m & -\lambda & -\mu \end{array}\right)
\left(\begin{array}{ccc} \bar l(-\bar\kappa) & 1/2 & \bar j \\ 
\bar m & -\lambda & -\bar\mu \end{array}\right)
\; .
\label{winkel-13}
\end{eqnarray}
This can be rewritten as
\begin{eqnarray}
\lefteqn{
   B_{ \kappa\mu ; \bar\kappa\bar\mu }
   = (-1)^{\bar\mu+1/2} \frac{1}{\sqrt{4\pi}} \,
   \exp(\iunit[\delta(-W,\kappa)-\bar\delta(-W,\bar\kappa)]) 
}
\nonumber
\\
   & \times &
   \sqrt{2j+1} \, \sqrt{2\bar j+1}
\nonumber
\\
   & \times &
   \sum\limits_{I , \alpha} \sqrt{2I+1} \, Y_{I\alpha}(\Omega_{p})
   \left(\begin{array}{ccc} j & \bar j & I \\ 1/2 & -1/2 & 0 
   \end{array}\right)
   \left(\begin{array}{ccc} \bar j & j & I \\ -\bar\mu & \mu & \alpha 
   \end{array}\right)
   \quad.
\label{koeff-b}
\end{eqnarray}
Furthermore we obtain the parity and angular momentum selection rules:
$$l(-\kappa)+\bar l(-\bar\kappa) +I = 0 \mbox{ mod } 2\quad,$$
$$|l(-\kappa) - \bar l(-\bar\kappa)| \leq I 
\leq l(-\kappa) + \bar l(-\bar\kappa) \quad.$$

\newpage
\section*{Figure captions:}
{\bf Figure 1:}
Opening-angle distribution of electron-positron pairs emitted by
IPC of randomly oriented uranium-like nuclei 
a) for an E1 transition, b) for a M1 transition. 
The transition energy amounts in both cases to 2 MeV, the
kinetic positron energy was taken to be 800 keV.
The full lines correspond to the DWBA calculations,
dotted lines display the outcome of 
the Born approximation. The effect of the finite
nuclear extension in the case of M1 transitions can be
deduced from the dashed line.

{\bf Figure 2:}
a) The coordinate system which is chosen to describe the alignment of
the Coulomb excited nuclei in heavy-ion collisions. $\vartheta$ denotes
here the scattering angle of the projectile in the lab system.
b) Definition of the angles which are used to describe the
directions of electron and positron in space. $\Theta$ is the opening
angle of the electron-positron pair, $\vartheta$ denotes the 
polar angle of the positron with respect to the quantization axis, and
the dihedral angle $\delta$ indicates the angle about which the
electron-positron emission plane is rotated around the positron axis.

{\bf Figure 3:}
Angular correlation of electron-positron pairs in E1-IPC 
for a transition from a
$1^{-}$ to a $0^{+}$ state, assuming oblate alignment
of uranium-like nuclei in the initial state. 
The conversion probability is plotted versus
the opening angle of the emitted lepton pair for various polar angles
of the positron with respect to the $z$ axis. The transition energy
amounts to $\omega = 1800$ keV, 
the kinetic positron energy to $E = 700$ keV.
The dihedral angle $\delta$ was fixed to $0^\circ$.

{\bf Figure 4:}
Angular correlation of electron-positron pairs in E1-IPC for
the same transition and energies as in Fig.~3. The pair conversion
probability is plotted for fixed polar angle $\vartheta=90^\circ$ and 
several values of the dihedral angle $\delta$.

{\bf Figure 5:}
Polar angle distribution assuming E1, E2, and E3 transitions to the
$0^+$ ground state of
a nucleus showing oblate alignment. The transition energy amounts to
$1800$ keV, the kinetic positron energy was fixed to $700$ keV.

{\bf Figure 6:}
Polar angle distribution assuming an E1 transition to the
$0^+$ ground state of a nucleus with oblate alignment, 
prolate alignment, and polarization. The transition energy amounts
to $1800$ keV, the kinetic positron energy to $700$ keV.

\newpage
\begin{tabular}{c}
\epsfysize=10cm\epsffile{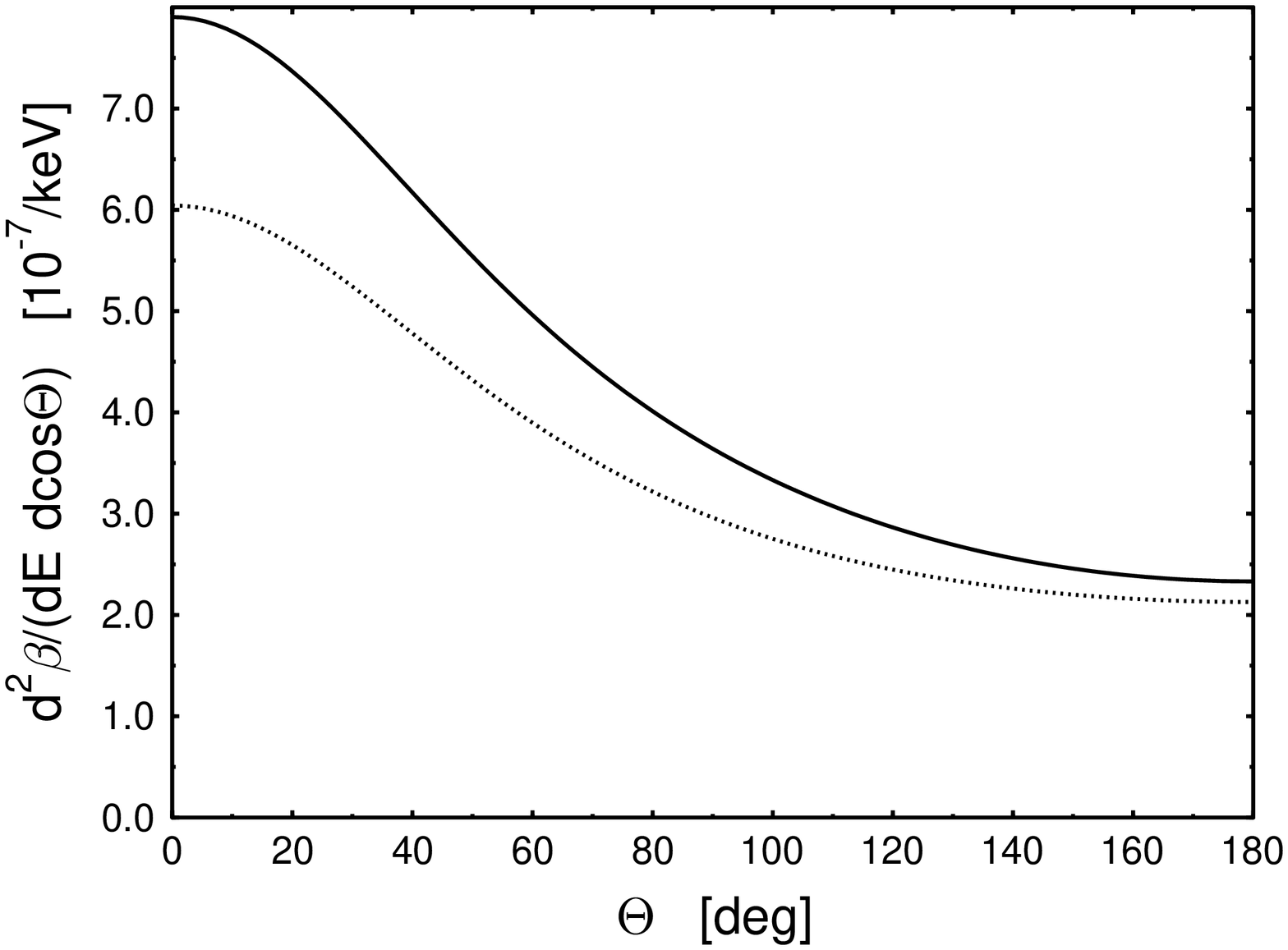} \\
{\bf Figure 1a}
\\
\epsfysize=10cm\epsffile{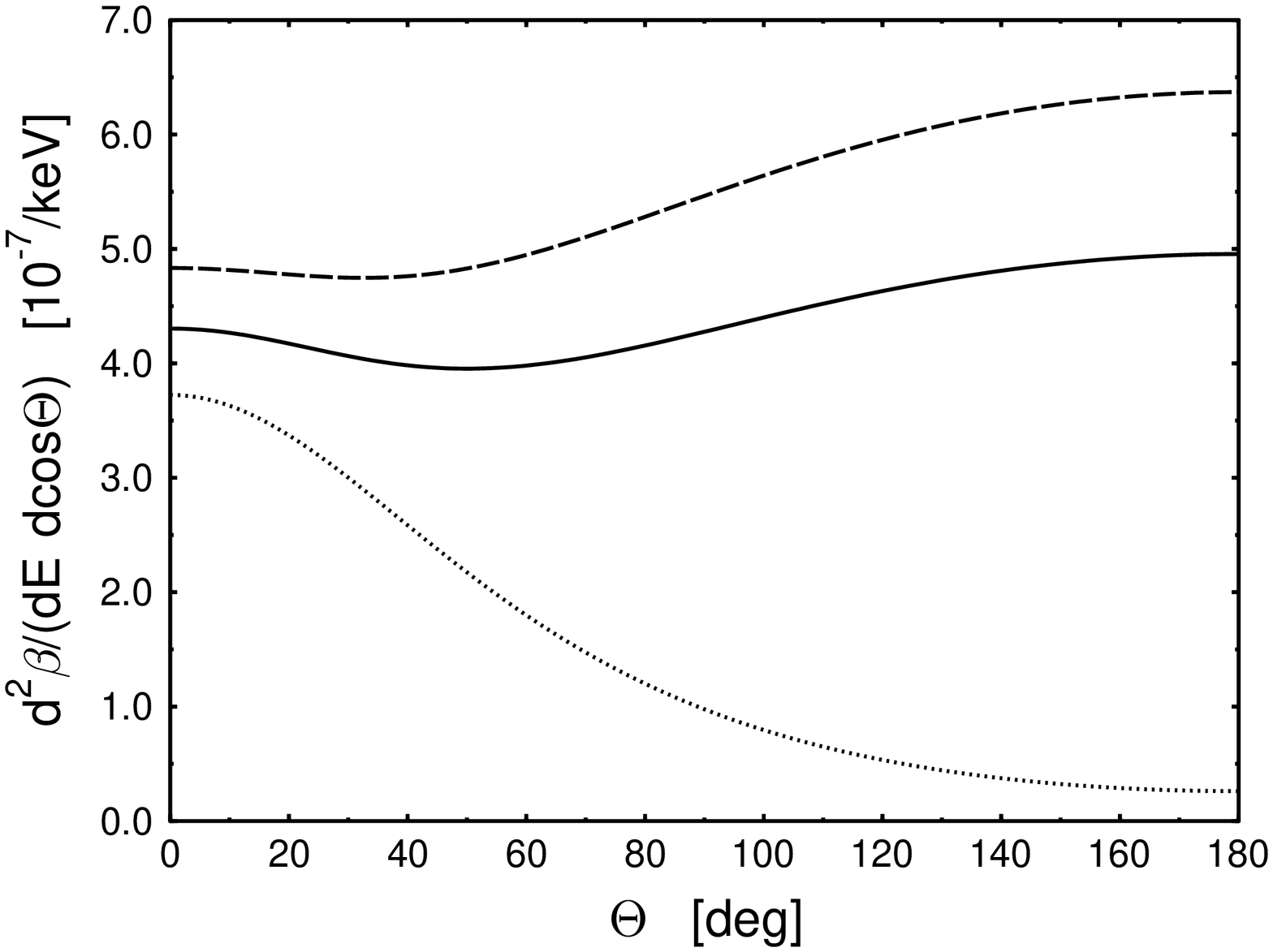} \\
{\bf Figure 1b}
\end{tabular}

\newpage
\begin{tabular}{c}
\epsfysize=10cm\epsffile{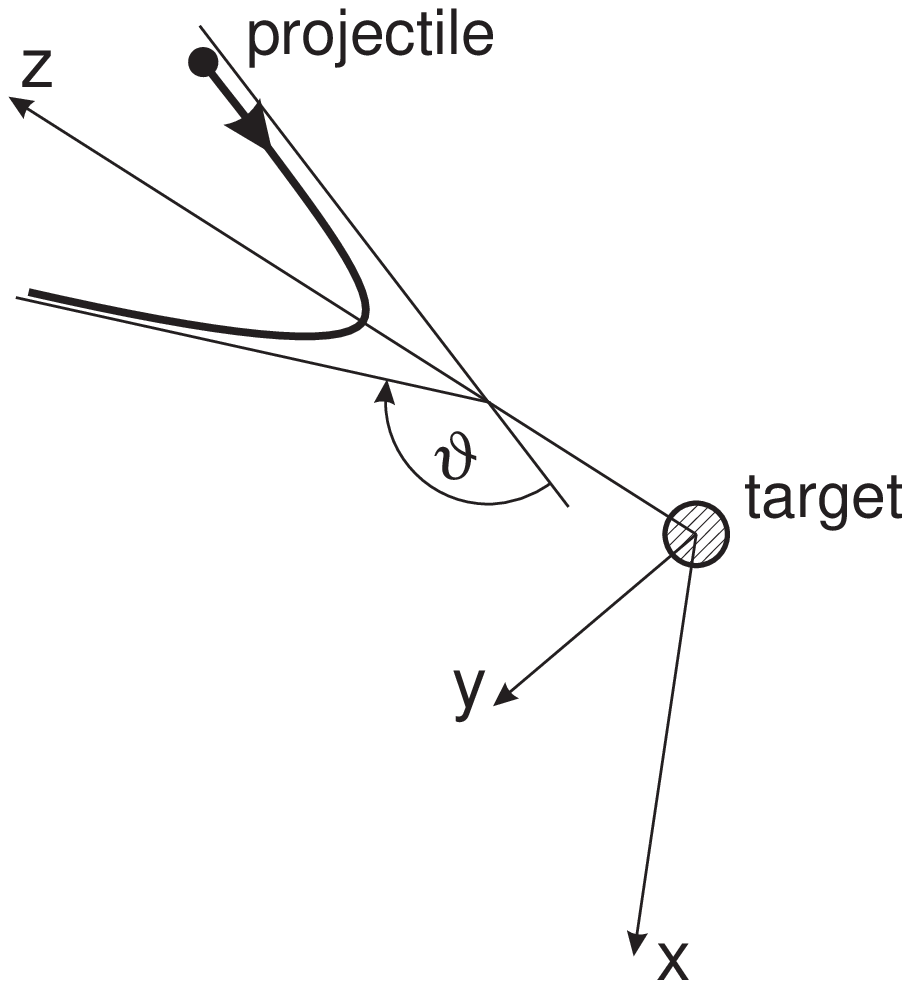} \\[1cm]
{\bf Figure 2a} 
\\[2cm]
\epsfysize=7cm\epsffile{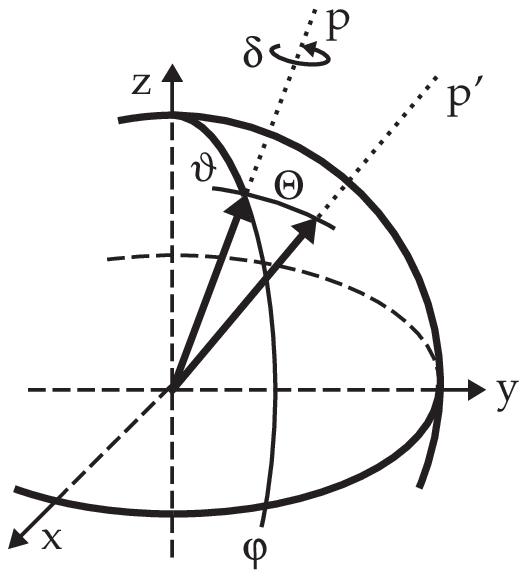} \\[1cm]
{\bf Figure 2b}
\end{tabular}


\newpage
\centerline{\hbox{
\epsfxsize=16cm\epsffile{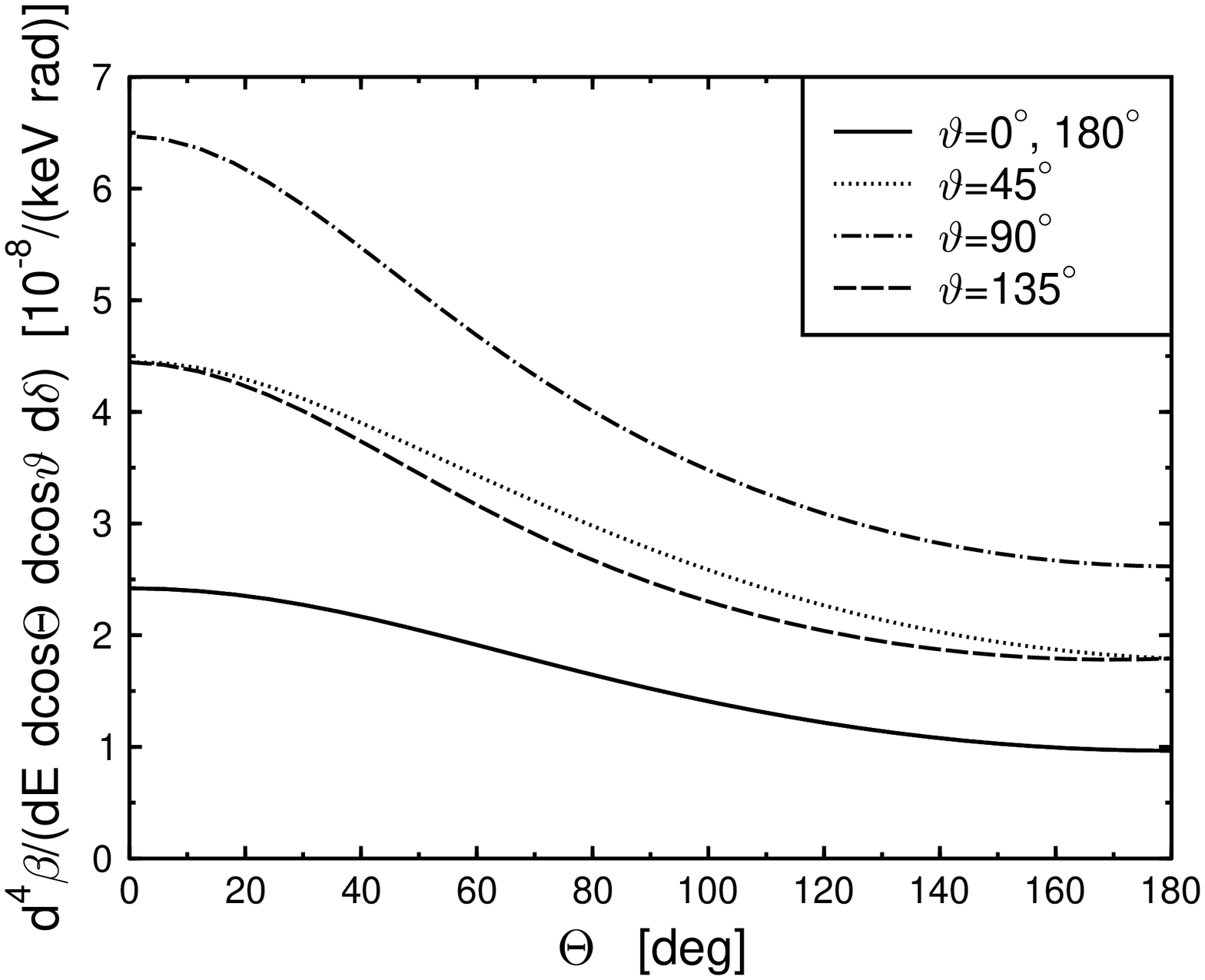}}}

\centerline{{\bf Figure 3}}

%

\newpage
\centerline{\hbox{
\epsfxsize=16cm\epsffile{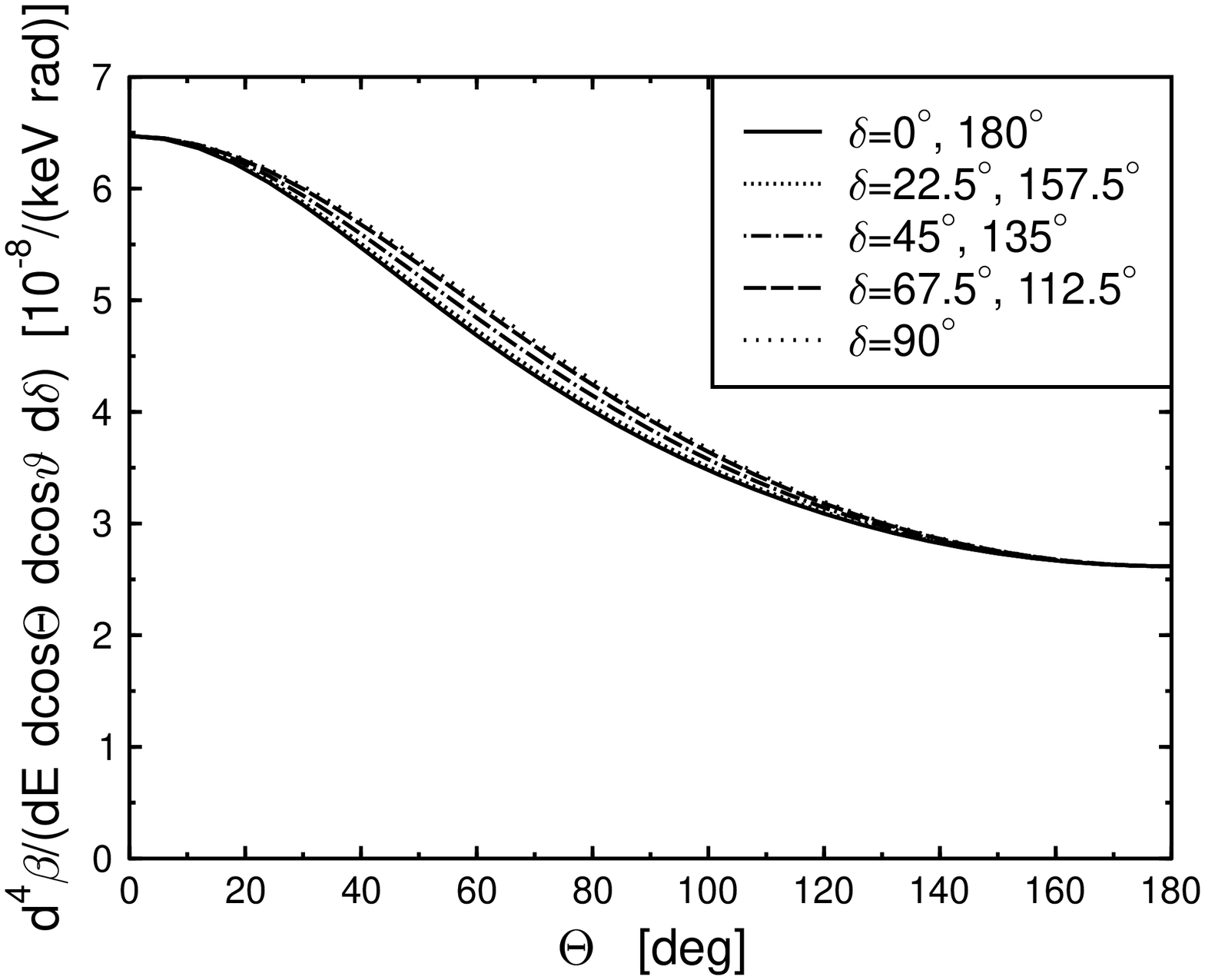}}}
\centerline{{\bf Figure 4}}

\newpage
\centerline{\hbox{
\epsfxsize=16cm\epsffile{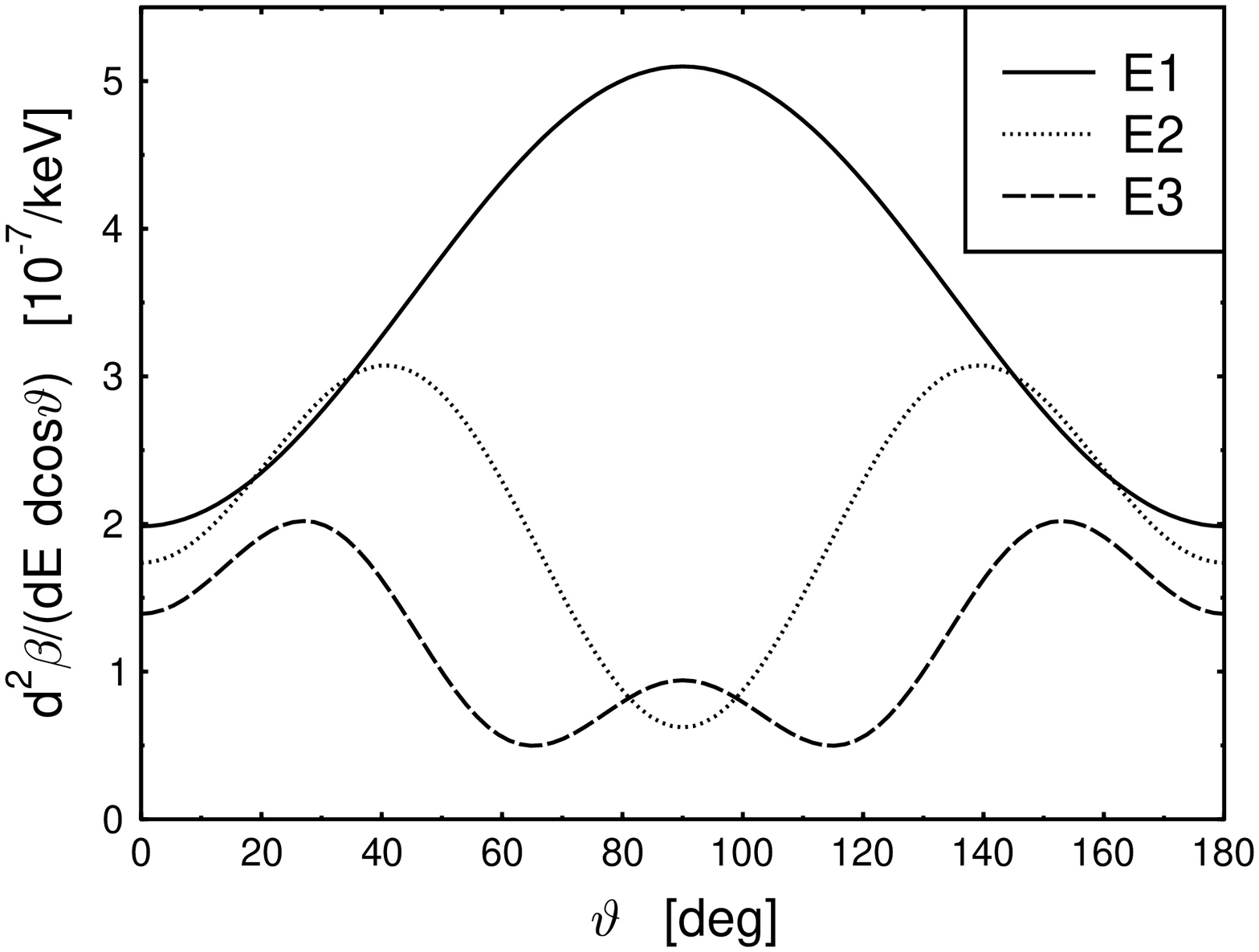}}}
\centerline{{\bf Figure 5}}

\newpage
\centerline{\hbox{
\epsfxsize=16cm\epsffile{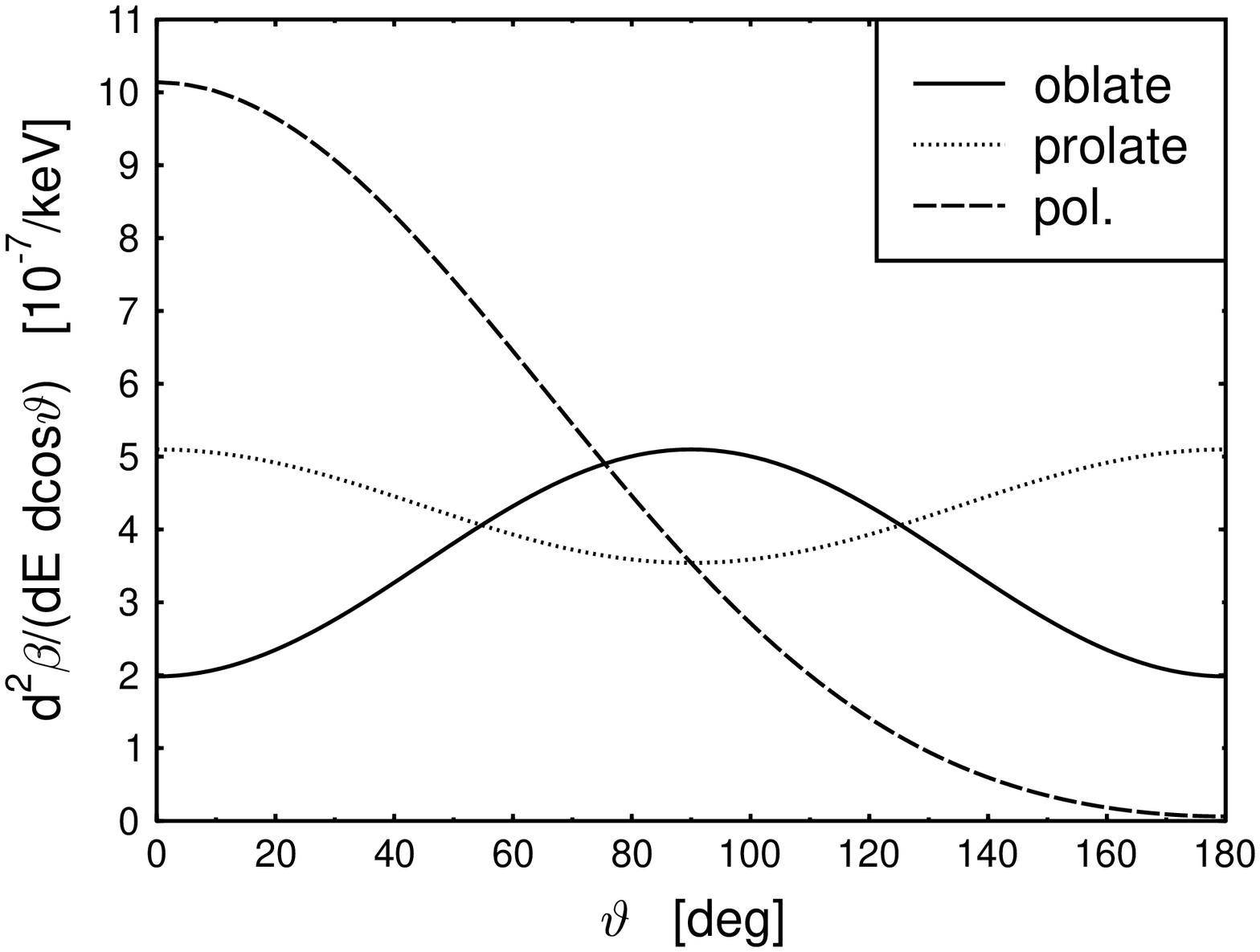}}}
\centerline{{\bf Figure 6}}

\end{document}